\documentstyle[preprint,prl,aps]{revtex}

\def\s1w{$\sigma_1 (\omega)$}
\def\cm-1{cm$^{-1}$}

\begin{document}

\draft

\title{X-ray absorption and optical spectroscopy studies
of (Mg$_{1-x}$Al$_x$)B$_2$}

\author{H. D. Yang$^{1}$, H. L. Liu$^{2}$, J.-Y. Lin$^{3}$,
M. X. Kuo$^{2}$, P. L. Ho$^{1}$, J. M. Chen$^{4}$, C. U.
Jung$^{5}$, Min-Seok Park$^{5}$, and Sung-Ik Lee$^{5}$}

\address{
$^{1}$Department of Physics, National Sun Yat-Sen University,
Kaohsiung 804, Taiwan ROC\\
$^{2}$Department of Physics, National Taiwan Normal University, 88, Sec. 4, Ting-Chou Road, Taipei 116, Taiwan ROC\\
$^{3}$Institute of Physics, National Chiao Tung University,
Hsinchu 300, Taiwan ROC\\
$^{4}$Synchrotron Radiation Research Center (SRRC), Hsinchu 300,
Taiwan ROC\\
$^{5}$National Creative Research Initiative center for
Superconductivity and Department of Physics, Pohang University of
Science and Technology, Pohang 790-784, Republic of Korea }
\date{\today}
\maketitle

\begin{abstract}
X-ray absorption spectroscopy and optical reflectance measurements
have been carried out to elucidate the evolution of the electronic
structure in (Mg$_{1-x}$Al$_{x}$)B$_{2}$ for $\emph{x}$ = 0.0,
0.1, 0.2, 0.3, and 0.4. The important role of B 2$\emph{p}$
$\sigma$ hole states to superconductivity has been identified, and
the decrease in the hole carrier number is $\emph{quantitatively}$
determined. The rate of the decrease in the hole concentration
agree well with the theoretical calculations. On the other hand,
while the evolution of the electronic structure is gradual through
the doping range, $T_c$ suppression is most significant at
$\emph{x}$ = 0.4. These results suggest that the superstructure in
(Mg$_{1-x}$Al$_{x}$)B$_{2}$, in addition to the $\sigma$ holes,
can affect the lattice dynamics and contributes to the $T_c$
suppression effect. Other possible explanations like the
topological change of the $\sigma$ band Fermi surface are also
discussed.
\end{abstract}


\narrowtext

Within two years of the discovery of superconductivity in
MgB$_{2}$\cite{Nagamatsu01}, intensive studies have led to
tremendous understanding of this new and unique intermetallic
superconductor. The main frame has been set both by theory and
experiments. MgB$_{2}$ is a phonon-mediated strong coupling
superconductor. The two-dimensional B 2$\emph{p}$ $\sigma$ holes
play a crucial role in superconductivity of
MgB$_{2}$\cite{Liu01,Singh01,Suzuki01,Choi}. In contrast to
classical $\emph{s}$-wave superconductors, MgB$_{2}$ has multiple
superconducting energy gaps though it is fully gapped
\cite{Liu01,Choi,Yang01,Bouquet01,Takahasi01,Tsuda01,Chen01}.
However, there are issues on MgB$_{2}$ remaining somewhat open
yet. One example is the Al doping effects on $T_c$ of MgB$_{2}$.
It is established that the Al doping leads to $T_c$ suppression in
MgB$_{2}$. Al doping also shortens the \emph{c} axis in the
layered hexagonal structure. Furthermore, the Al-layer ordering in
(Mg$_{1-x}$Al$_{x}$)B$_{2}$ was reported \cite{Slusky01,Li02}.
Since superconductivity in MgB$_{2}$ is of phonon origin, the
interplay of the lattice dynamics and the evolution of the
electronic structure in (Mg$_{1-x}$Al$_{x}$)B$_{2}$ is of great
interest. However, compared to the detailed studies of phonon
spectrum in (Mg$_{1-x}$Al$_{x}$)B$_{2}$\cite{Renker02}, the Al
doping effects on the electronic structure have not been
thoroughly explored yet. In this paper, we report detailed x-ray
absorption and optical reflectance measurements in
(Mg$_{1-x}$Al$_{x}$)B$_{2}$ for \emph{x} = 0.0, 0.1, 0.2, 0.3, and
0.4. Comparisons between the present results and other theoretical
and experimental works provide deeper insight into the electronic
structure and $T_c$ suppression mechanism in
(Mg$_{1-x}$Al$_{x}$)B$_{2}$.

To prepare (Mg$_{1-x}$Al$_{x}$)B$_{2}$, the stoichiometric mixture
of Mg, $^{11}$B, and Al powder (Alfa Aesar) was ground softly for
an hour. Resultant powder was palletized and wrapped by a Ta foil.
Then it was put into a high pressure cell. This whole process was
performed in an inert Ar gas. A 12 mm cubic multi-anvil system was
used for a high pressure synthesis. The cell was heated up to 950
$^{o}$C and maintained 950 $^{o}$C for 2 hours. Then it was
quenched to room temperature. Details of high-pressure synthesis
will be found elsewhere\cite{Jung01,Jung02}. The lattice
parameters $\emph{a}$ and $\emph{c}$, shown in the inset of
Fig.~1, were obtained from x-ray diffraction measurements. Al
doping led to an obvious decrease in $\emph{c}$ due to the smaller
ionic radius of Al$^{3+}$than that of Mg$^{2+}$, while the effect
on $\emph{a}$ was relatively insensitive. The diffraction pattern
of the $\emph{x}$ = 0.2 sample showed the presence of two phases.
All the above x-ray diffraction results were consistent with those
in the literature \cite{Slusky01,Li02}. The resistivity $\rho$ was
measured by the four-probe method, as shown in Fig. 1. $T_c$ was
determined by the midpoint of the transition, and was consistent
with the magnetization $\emph{M}$ measurements. The sharp
transition both in $\rho$ and $\emph{M}$ manifested the good
quality of the samples. Moreover, $T_c$ suppression effects due to
Al doping are consistent with the reported
results\cite{Slusky01,Li02}.

X-ray absorption near edge structure (XANES) in fluorescence mode
is a powerful tool to investigate the unoccupied (hole) electronic
states in complex materials and is bulk sensitive. The B
$\emph{K}$-edge x-ray absorption spectra were carried out using
linear polarized synchrotron radiation from 6-m high-energy
spherical grating monochromator beamline located at SRRC in
Taiwan. Details of the measurements were described elsewhere
\cite{Chen97,Hong02}. Energy resolution of the monochromator is
set to be ~0.15 eV for the B $\emph{K}$-edge energy range. The
energy was calibrated as in Ref. ~\onlinecite{Nakamura01}. The
absorption spectra were normalized to the maximum of the peak
around 200 eV.

Near-normal optical reflectance spectra were taken at room
temperature on mechanically polished surfaces
of high density polycrystalline samples.
Middle infrared (600-3000 \cm-1) measurements were made
with a Perkin-Elmer 2000 spectrometer coupled with a FT-2R microscope,
while the spectra in the near-infrared to near-ultraviolet regions
(4000-55000 cm$^{-1}$) were collected on a
Perkin-Elmer Lambda-900 spectrometer.
The modulated light beam from the spectrometer was
focused onto either the sample or an Au (Al) reference mirror, and
the reflected beam was directed onto a detector appropriate for the
frequency range studied. The different sources and detectors used
in these studies provided substantial spectral overlap, and the
reflectance mismatch between adjacent spectral ranges was less
than 1 $\%$.

The B $\emph{K}$-edge XANES on  (Mg$_{1-x}$Al$_{x}$)B$_{2}$ was
studied and shown in Fig. 2. The peaks centered between 186.5 and
187 eV can be identified to be closely associated with B
2$\emph{p}$ $\sigma$ holes. Two additional peaks around
192$\sim$194 eV (not shown), probably due to either boron oxides
or resonances, were also observed in this work. These features do
not appear to affect the peak of B 2$\emph{p}$ $\sigma$ states,
and are not included in the following
discussions\cite{Nakamura01,Callcott01,Schuppler}.

The most noticeable change is the decrease in the intensity of the
pre-edge peak with increasing Al doping. This can be reasoned as
the electron doping effect with the substitution of Al$^{3+}$ for
Mg$^{2+}$. XANES thus directly verifies the crucial role of B
2$\emph{p}$ $\sigma$ holes together with $T_c$ suppression effects
due to Al doping. It is also noticed that the change of the
intensity is gradual, like the $\emph{x}$ dependence of
$T_c$($\emph{x}$), until $\emph{x}$ = 0.4. No dramatic change
between $\emph{x}$ = 0.0 and 0.1 was observed in the present work,
while a 65$\%$ drop in the spectral weight was reported in
Ref.~\onlinecite{Schuppler}. In general, the onset and the peak
energies both increase with Al doping as expected naively by the
concept of hole filling, while indicating that the rigid band
model can not be vigorously applied. Furthermore, the onset and
the peak energies actually decrease slightly from $\emph{x}$ = 0.0
to 0.1, which probably suggests a corresponding change of the core
level energy. This curious tendency appears to be genuine, since
it was also observed in another independent work \cite{Tsuei}.

To further quantify the decrease in the number of hole carriers,
the optical spectroscopy has been proved to be an effective tool for the
investigation of the doping-induced spectral weight\cite{Tanner98}.
The optical properties
({\it {i.e.}} the complex conductivity $\sigma(\omega) =
\sigma_1(\omega) + i\sigma_2(\omega)$ or dielectric function
$\epsilon(\omega)$ = 1 + ${4\pi i \sigma(\omega)} / \omega$) were
calculated from Kramers-Kronig analysis of the reflectance
data\cite{Wooten72}. To perform these transformations one needs to
extrapolate the reflectance at both low and high frequencies. At
low frequencies the extension was done by modeling the reflectance
using the Drude model and using the fitted results to extend the
reflectance below the lowest frequency measured in the experiment.
The high-frequency extrapolations were done by using a weak power
law dependence, R $\sim$ $\omega^{-s}$ with s $\sim$ 1--2.

Similar optical experiments has been performed on
MgB$_{2}$\cite{Gorshunov01}. In this work, the detailed optical
measurements on (Mg$_{1-x}$Al$_{x}$)B$_{2}$ are reported. Figure~3
shows the room-temperature real part of the optical conductivity
$\sigma_1(\omega)$ as a function of Al doping, obtained from a
Kramers-Kronig transformation of the measured reflectance data.
For the x = 0.0 sample, the optical conductivity can be described
in general terms as (i) coherent response of itinerant charge
carries at zero frequency; (ii) an overdamped mid-infrared
component around 3500 cm$^{-1}$; and (iii) two interband
transitions near 17000 and 50000 cm$^{-1}$. We have tried to fit
the conductivity spectrum in the whole frequency range with a
Drude part and three Lorentz oscillators. The Drude plasma
frequency of the carriers $\omega_{pD}$ and their scattering rate
$1/\tau_D$ are 27000 and 1000 cm$^{-1}$, respectively. The
estimated Drude resistivity [$\rho_{\rm Drude}$ =
($\omega_{pD}^2$$\tau_D$/60)$^{-1}$, in unit $\Omega$-cm] is
8$\times$10$^{-5}$$\Omega$-cm, in reasonable agreement with the
transport results. As Al doping proceeds, the far-infrared
conductivity is decreasing, while the oscillator strength of the
mid-infrared absorption is nearly independent of doping.

An attempt in separating the zero-frequency absorption channel
from the mid-infrared absorption sometimes produced ambiguous
results. This is the case that there has been much discussion over
the one-component and the two-component pictures to describe the
optical conducitivity of high-T$_c$ cuprates\cite{Tanner92}. The
doping dependence of the low-frequency optical conductivity of
these samples can also be summarized by plotting the integrated
spectral weight in the conductivity\cite{Wooten72},
$$
{N_{\rm eff}(\omega)} =
{{ {2m_0 V_{\rm cell}} \over {\pi e^2}} \int\limits_0^\omega {\sigma_1(\omega^{'})}
d\omega^{'}},
\eqno(2)
$$
where $m_0$ is taken as the free-electron mass, and $V_{\rm cell}$
is the unit cell volume. $N_{\rm eff}(\omega)$ is proportional to
the number of carriers participating in the optical absorption up
to a certain cutoff frequency $\omega$, and has the dimension of
frequency squared. Integration of the conductivity up to $\omega$
= 8000 cm$^{-1}$- the frequency at which we observe a clear onset
of interband transitions-provides only 30{\%} of the spectral
weight we measure when $\omega$ is extended up to our experiment
limit of 52000 cm$^{-1}$. Fig.~4 shows the quantity $N_{\rm eff}$
($\omega$ = 8000 cm$^{-1}$) plotted as a function of $T_c$,
illustrating the correlation between the number of carriers and
$T_c$ in the (Mg$_{1-x}$Al$_x$)B$_2$ systems. It is interesting to
note that the spectral weight corresponds to an effective number
of carriers of $\sim$ 0.48 and 0.16 for the x = 0.0 and 0.4
samples, suggesting that each Al removes approximately one
carrier. From Fig.~4, it is intriguing to note that $T_c$ changes
accordingly with $N_{\rm eff}$ until a more significant drop in
$T_c$ occurs at $\emph{x}$ = 0.4. This implies that an additional
effect other than $N_{\rm eff}$ begins to play a role in the
determination of $T_c$ at the composition of $\emph{x}$ = 0.4.
Another intriguing plot is $N_{\rm eff}$ vs $\emph{x}$ as in
Fig.~5. It unambiguously demonstrates $\emph{no}$ abrupt change of
$N_{\rm eff}$ at either $\emph{x}$ = 0.1 or 0.4, in agreement with
XANES in Fig.~3. It has been argued that the evolution of the
electronic structure is responsible for the change of the lattice
dynamics in (Mg$_{1-x}$Al$_x$)B$_2$\cite{Renker02}. The results in
Figs. 4 and 5 further suggest that the phonon spectrum could be
affected by the superstructure, in addition to the factor of
$N_{\rm eff}$. Why this superstructure effect takes place at
$\emph{x}$ = 0.4 rather than at smaller $\emph{x}$ is unknown.
Perhaps the composition of $\emph{x}$ = 0.4 approaches $\emph{x}$
= 0.5 close enough where the Al layer ordering is much preferred.
The formation of the superstructure has been explained recently by
a simple model, though with a too low instability temperature for
the phase separation\cite{barabash02}. The interplay of the
electronic structure, the lattice dynamics, and the superstructure
still remains unclear with respect to the details.

B 2$\emph{p}$ $\sigma$ holes in (Mg$_{1-x}$Al$_x$)B$_2$ were
actually theoretically investigated\cite{Suzuki01}. Figure~5
provides an excellent stage to compare between theoretical
calculations and experimental results. Though the absolute value
of the carrier number is difficult to compare, the comparison of
the relative change of the carrier number due to Al doping can be
made. Theoretical calculations predict $N_{\rm eff}$(\emph{x})/
$N_{\rm eff}$(0)=1-1.75$\emph{x}$\cite{Suzuki01}, which amazingly
agrees with the experimental results in Fig.~5. It is noted that
the solid line in Fig.~5 represents the theoretical prediction
$\emph{with no fitting parameter}$. This astonishing agreement
deserves further consideration. In the calculations of Ref.
~\onlinecite{Suzuki01}, doped Al atoms were assumed to be
distributed in Mg planes, and no superstructure effects were
included. It seems that the occurrence of the superstructure has
no effect on $N_{\rm eff}$ while it affects the lattice dynamics
which contributes to further $T_c$ suppression for
$\emph{x}\geq$0.4. This scenario is qualitatively in accord with
the experimental results\cite{Renker02}. It is also likely that
the coupling between $\sigma$ and $\pi$ bands has to be included
to account for the correct $T_c$ suppression
effects\cite{barabash02}. Furthermore, another first principle
calculation work (without considering the superstructure effects)
suggests that an abrupt topological change in the $\sigma$ Fermi
surface between $\emph{x}$=0.3 and 0.4 \cite{pena02}. How this
electronic structure change reconciles with the observed smooth
change of XANES and $N_{\rm eff}$(\emph{x}), and how it is related
to the evolution of the lattice dynamics and the $T_c$ suppression
remain largely unknown. Certainly further theoretical studies as
in Ref. ~\onlinecite{Suzuki01} but including the interplay of all
the electronic structure, the lattice dynamics, and the
superstructure effects are indispensable.

To conclude, spectroscopy studies of (Mg$_{1-x}$Al$_x$)B$_2$ by
XANES and optical reflectance measurements not only identify the
importance of B 2$\emph{p}$ $\sigma$ hole states to
superconductivity, but also quantify the changes of the carrier
number due to Al doping. The quantitative decreasing rate of the
$\sigma$ hole number agrees well with the theoretical
calculations. No abrupt change of the carrier number was observed
for $\emph{x}$ = 0.0 to 0.4. This smooth change is difficult to
reconcile with the large $T_c$ suppression at $\emph{x}$=0.4.
While either the superstructure or the topological change of the
electronic structure in (Mg$_{1-x}$Al$_x$)B$_2$ could be
responsible for this anomaly, the experimental results of the
electronic structure evolution indicate that a full understanding
of the large $T_c$ suppression at $\emph{x}$ = 0.4 is still
desirable.

\acknowledgements

We are grateful to B. Renker and R. Heid for inspiring discussions
in MOS2002 conference. This work was supported by the National
Science Council of the Republic of China under Grant Nos. NSC
91-2112-M-003-021, NSC91-2112-M-110-005 and NSC91-2112-M-009-046.

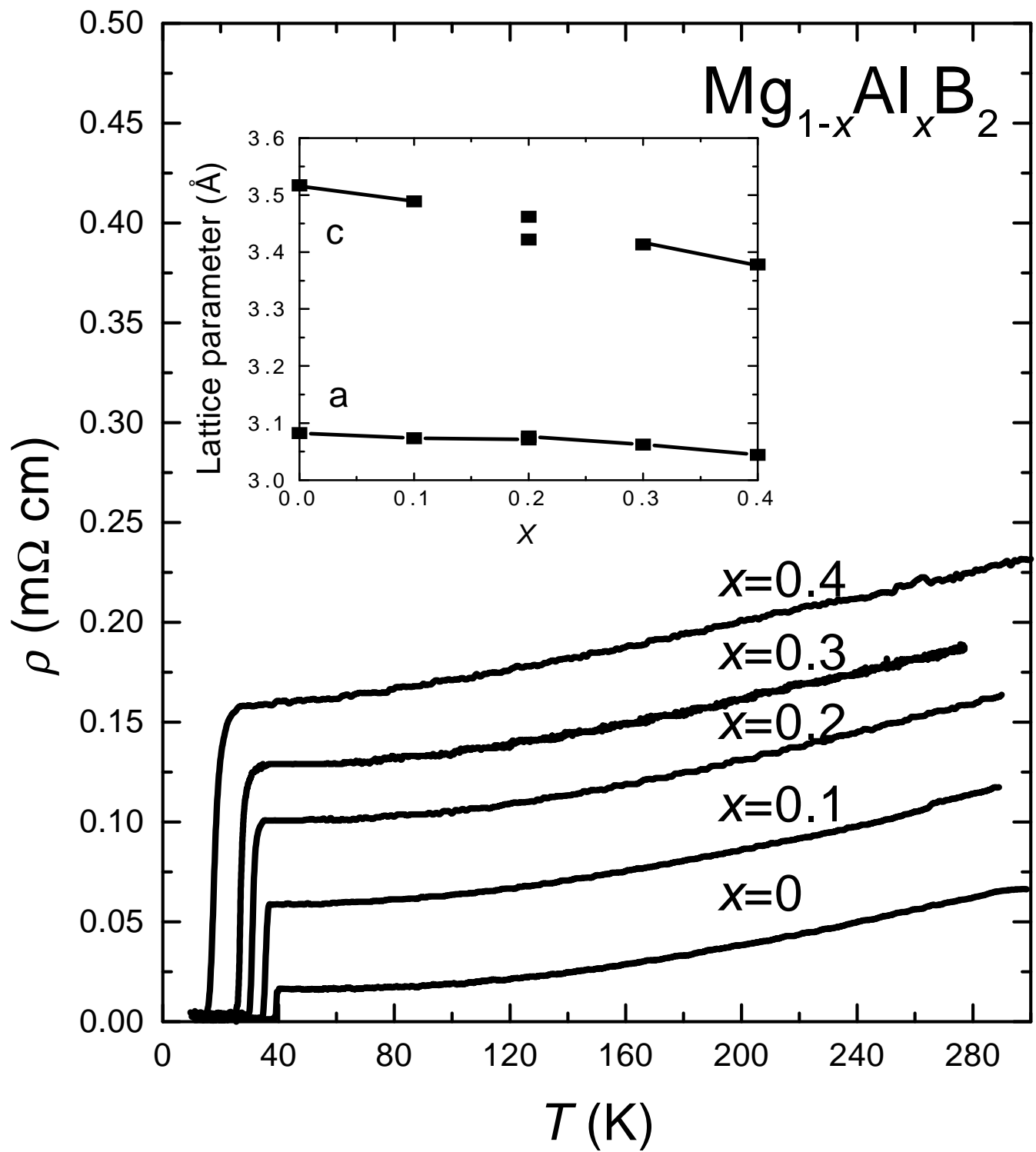
\begin{figure}
\caption{Resistivity $\rho$($\emph{T}$) of
(Mg$_{1-x}$Al$_x$)B$_2$. Inset: lattice parameters of
(Mg$_{1-x}$Al$_x$)B$_2$.}\label{Fig.1}
\end{figure}

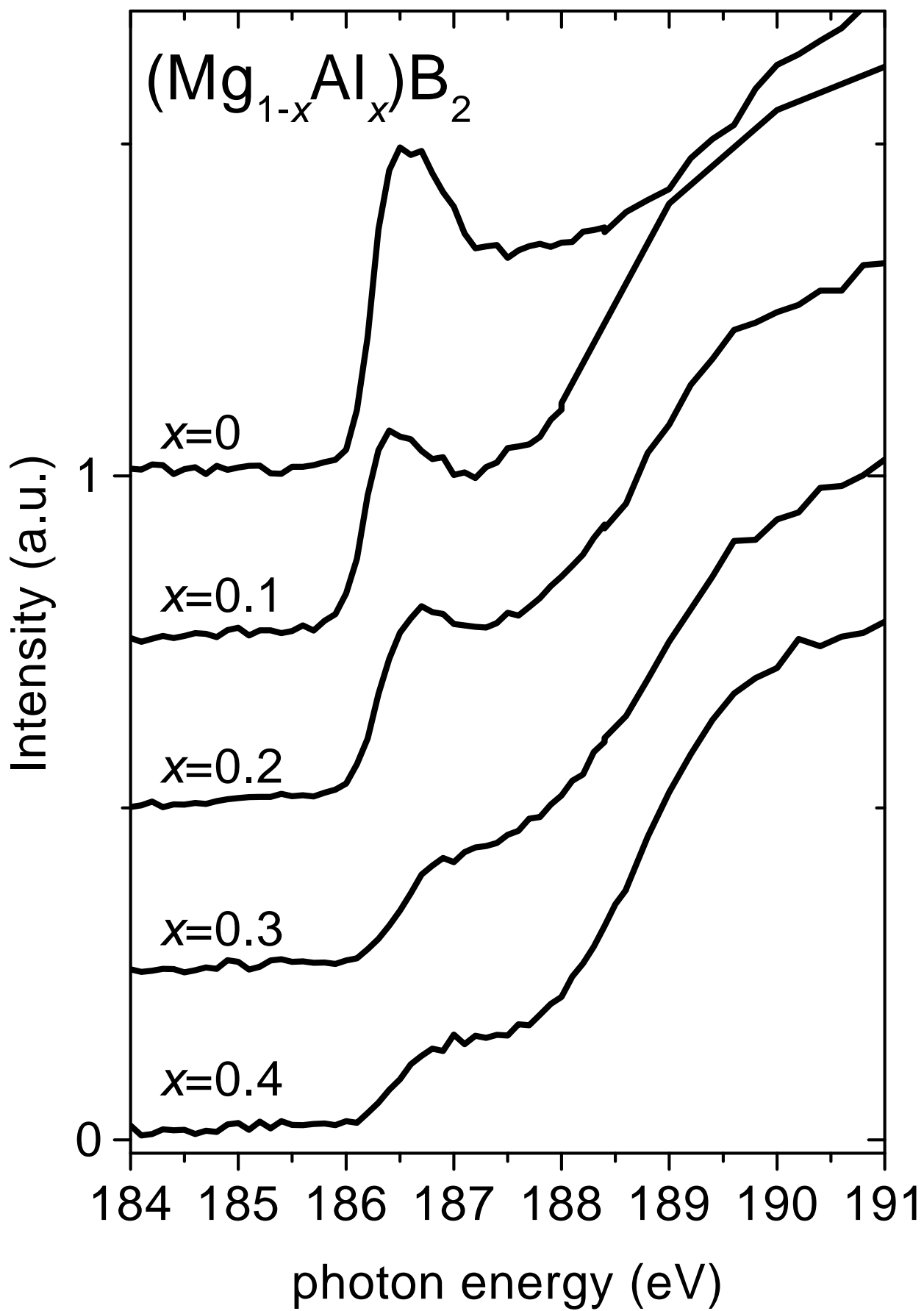
\begin{figure}
\caption{B $\emph{K}$-edge XANES of (Mg$_{1-x}$Al$_x$)B$_2$ for
$\emph{x}$=0 to 0.4. The pre-edge peak is associated with the B
2$\emph{p}$ $\sigma$ hole states.}\label{Fig.2}
\end{figure}

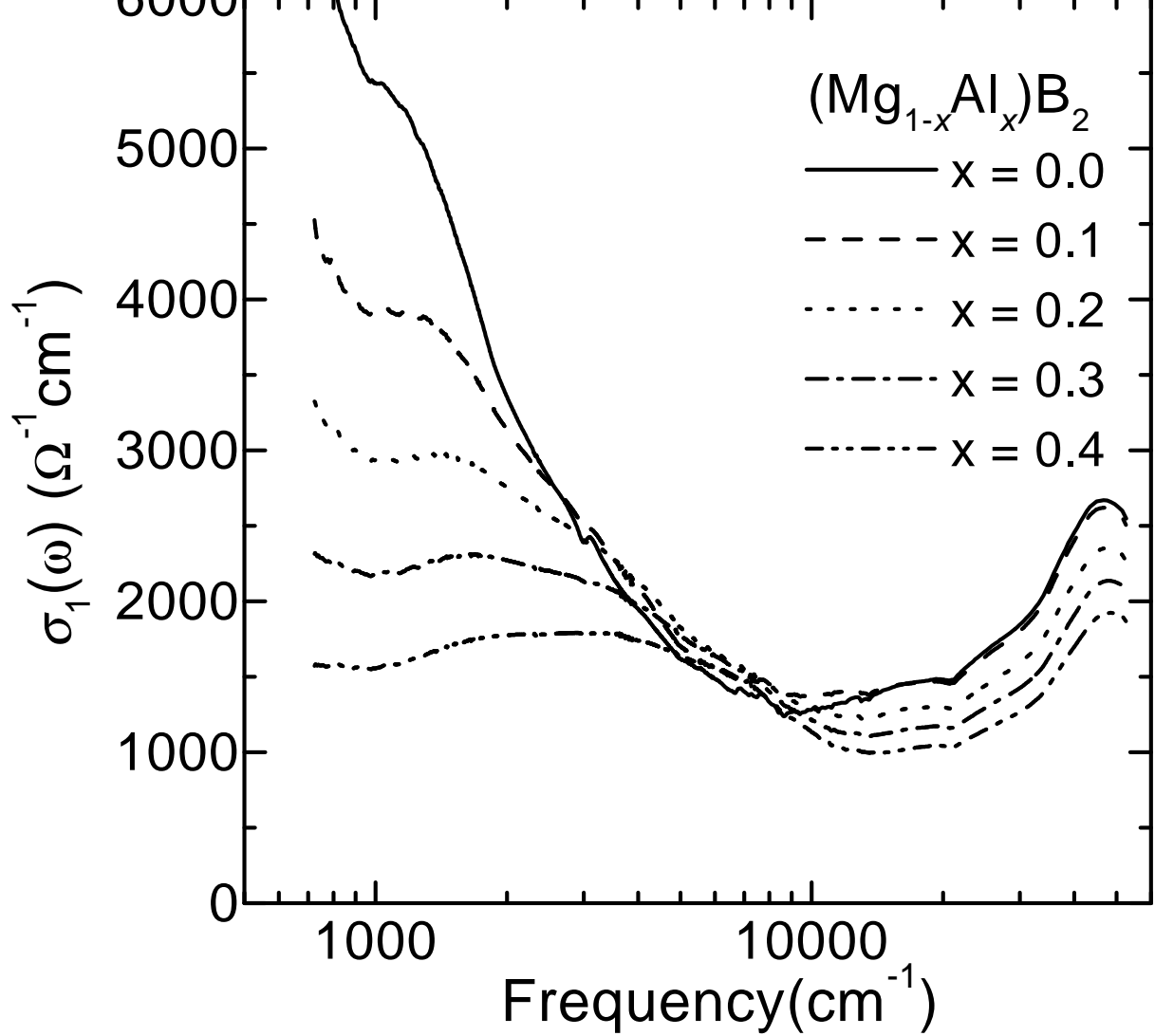
\begin{figure}
\caption{Room-temperature optical conductivity spectra of
(Mg$_{1-x}$Al$_x$)B$_2$. } \label{Fig.3}
\end{figure}

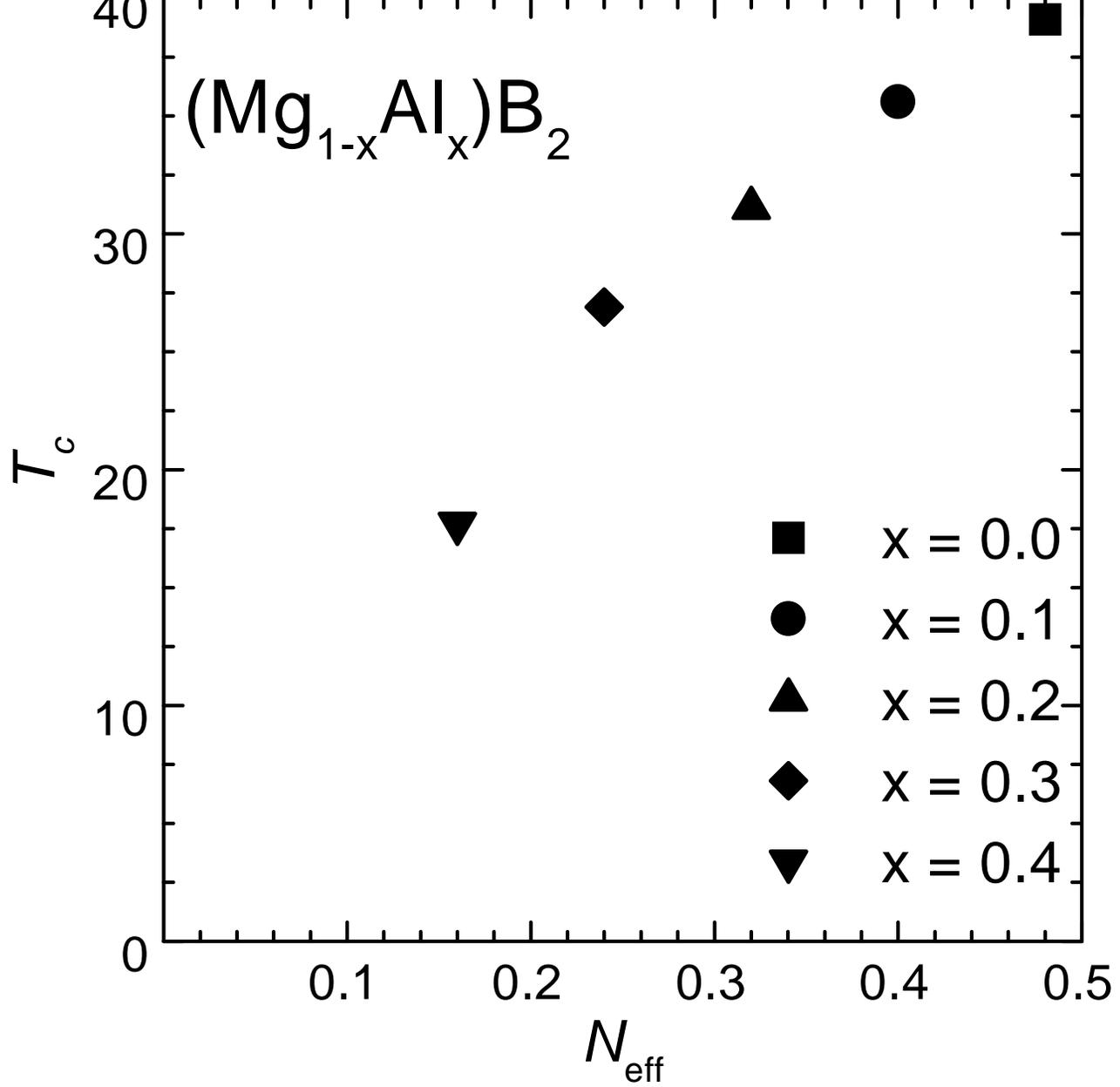
\begin{figure}
\caption{The effective number of carriers $N_{\rm eff}$ integrated
up to 8000 cm$^{-1}$ vs $T_c$ } \label{Fig.4}
\end{figure}

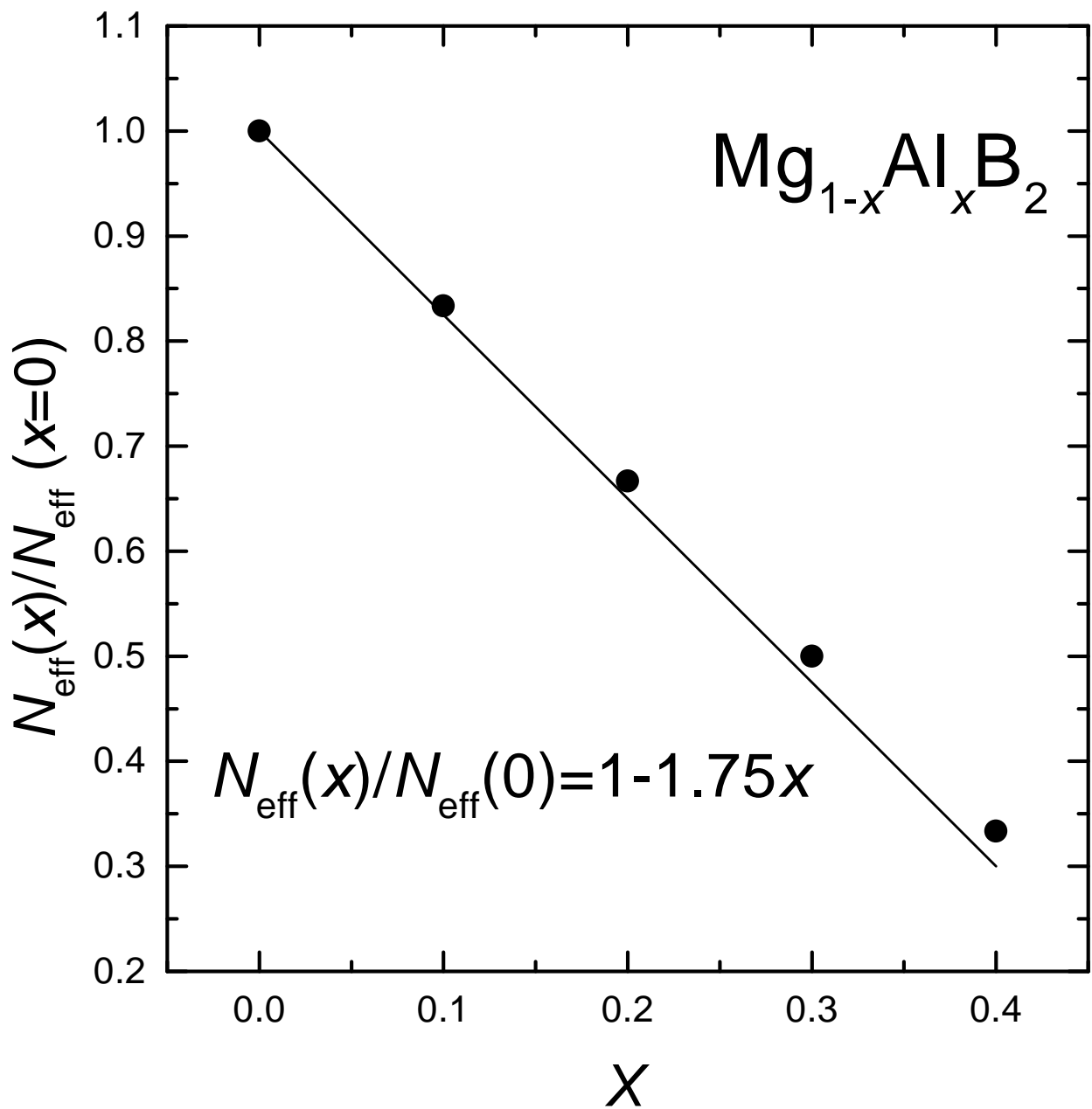
\begin{figure}
\caption{$N_{\rm eff}$ vs $\emph{x}$. The experimental results are
shown as the solid circles. The theoretical prediction is taken
from Ref. 4 and denoted as the solid line. No fitting parameter is
adjusted.} \label{Fig.5}
\end{figure}


\begin{references}

\bibitem{Nagamatsu01}J. Nagamatsu, N. Nakagawa, T. Muranaka, Y. Zenitani, and J. Akimitsu, Nature $\textbf{410}$, 63
(2001).
\bibitem{Liu01}A. Y. Liu, I. I. Mazin, and J. Kortus,
Phys. Rev. Lett. $\textbf{87}$, 087005 (2001).
\bibitem{Singh01}P. P. Singh, Phys. Rev. Lett. $\textbf{87}$, 087004 (2001).
\bibitem{Suzuki01}S. Suzuki, S. Higai, and K. Nakao, J. Phys. Soc. Jpn. $\textbf{70}$, 1206 (2001).
\bibitem{Choi}H. J. Choi, D. Roundy, H. Sun, M. L. Cohen, and S.
G. Louie, Nature $\textbf{418}$, 758 (2002).
\bibitem{Yang01}H. D. Yang, J.-Y. Lin, H. H. Li, F. H. Hsu, C.-J. Liu, S.-C. Li, R.-C. Yu and C.-Q. Jin, Phys. Rev. Lett. $\textbf{87}$, 167003 (2001).
\bibitem{Bouquet01}F. Bouquet, R. A. Fisher, N. E. Phillips, D. G. Hinks, and J. D. Jorgensen,Phys. Rev. Lett. $\textbf{87}$, 047001 (2001).
\bibitem{Takahasi01}T. Takahashi, T. Sato, S. Souma, T. Muranaka, and J. Akimitsu, Phys. Rev. Lett. $\textbf{86}$, 4915 (2001).
\bibitem{Tsuda01}S. Tsuda, T. Yokoya, T. Kiss, Y. Takano, K. Togano, H. Kito, H. Ihara, and S. Shin, Phys. Rev. Lett.
$\textbf{87}$, 177006 (2001).
\bibitem{Chen01}X. K. Chen, M. J. Konstantinovi$\acute{c}$, J. C. Irwin, D. D. Lawrie, and J. P. Franck, Phys. Rev. Lett. $\textbf{87}$, 157002 (2001).
\bibitem{Slusky01}J. S. Slusky, N. Rogado, K. A. Regan, M. A.
Hayward, P. Khalifah, T. He, K. Inumaru, S. Loureiro, M. K. Hass,
H. W. Zandbergen, and R. J. Cava, Nature $\textbf{410}$, 343
(2001).
\bibitem{Li02}J. Q. Li, L. Li, F. M. Liu, C. Dong, J. Y. Xiang,
and Z. X. Zhao, Phys. Rev. B $\textbf{65}$, 132505 (2002).
\bibitem{Renker02}B. Renker, K. B. Bohnen, R. Heid, D. Ernst, H.
Schober, M. Koza, P. Adelmann,  P. Schweiss, and Th. Wolf, Phys.
Rev. Lett. $\textbf{88}$, 067001 (2002), and references therein.
\bibitem{Jung01}C. U. Jung, Min-Seok Park, W. N. Kang, Mun-Seog Kim, Kijoon H. P. Kim, S. Y. Lee, and Sung-Ik Lee, Appl. Phys. Lett. $\textbf{78}$, 4157
(2001).
\bibitem{Jung02}C. U. Jung, Heon-Jung Kim, Min-Seok Park, Mun-Seog Kim, J. Y. Kim, Zhonglian Du, Sung-Ik Lee, K. H. Kim, J. B. Betts, M. Jaime, A. H. Lacerda, and G. S. Boebinger,
Physica C $\textbf{377}$, 21 (2002).
\bibitem{Chen97}J. M. Chen, R. S. Liu, J. G. Lin, C. Y. Huang, and J. C. Ho, Phys. Rev. B $\textbf{55}$, 14586 (1997).
\bibitem{Hong02}I. P. Hong, J.-Y. Lin, J. M. Chem, S. Chatterjee. S. J. Liu, Y. S. Gou, and H. D. Yang, Europhys. Lett. $\textbf{58}$, 126 (2002).
\bibitem{Nakamura01}J. Nakamura, N. Yamada, K. Kuroki, T. A.
Callcott, D. L. Ederer, J. D. Denlinger, and R. C. C. Perera,
Phys. Rev. B $\textbf{64}$, 174504 (2001).
\bibitem{Callcott01}T. A. Callcott, L. Lin, G. T. Woods, G. P.
Zhang, J. R. Thompson, M. Paranthaman, and D. L. Ederer, Phys.
Rev. B $\textbf{64}$, 132504 (2001).
\bibitem{Schuppler}S. Schuppler, E. Pellegrin, N. N$\ddot{u}$ker, T.
Mizokawa, M. Merz, D. A. Arena, J. Dvorak, Y. U. Idzerda, D.-J.
Huang, C.-F. Cheng, K.-P. Bohnen, R. Heid, P. Schweiss, and Th.
Wolf, cond-mat/0205230.
\bibitem{Tsuei}K.-D. Tsuei,H.-J. Lin, L.-C. Lin, T.-Y. Hou, H.-H. Hsieh, C. T. Chen, N. L. Saini, A. Bianconi, and A. Saccone, Inter. J. Mod. Phys. B $\textbf{16}$, 1619 (2002).
\bibitem{Tanner98} D. B. Tanner, H. L. Liu, M. A. Quijada, A. M. Zibold,
H. Berger, R. J. Kelley, M. Onellion, F. C. Chou, D. C. Johnston,
J. P. Rice, D. M. Ginsberg, and J. T. Markert, Physica B
$\textbf{244}$, 1 (1998).
\bibitem{Wooten72} F. Wooten, in {\it
Optical Properties of Solids} (Academic, New York, 1972).
\bibitem{Gorshunov01}For example, B. Gorshunov, C. A. Kuntscher,
P. Haas, M. Dressel, F. P. Mena, A. B. Kuz'menko, D. van der
Marel, T. Muranaka, and J. Akimitsu, Eur. Phys. J. B
$\textbf{21}$, 159 (2001).
\bibitem{Tanner92} D. B. Tanner and T. Timusk, in {\it Physical Properties of
High Temperature Superconductors III}, edited by D. M. Ginsberg
(World Scientific, Singapoer, 1992), p. 363.
\bibitem{barabash02}S. V. Barabash and D. Stroud, Phys. Rev. B
$\textbf{66}$,012509 (2002).
\bibitem{pena02}O. de la Pe$\tilde{n}$a, A. Aguayo and R. de Coss, Phys. Rev. B
$\textbf{66}$,012511 (2002).

\end{references}
\end{document}